# FULL-DEPTH SNOW AVALANCHES

François Louchet

284 Chemin du Pré Roudon, 384210 St Martin d'Uriage (France)

(formerly at "Laboratoire de Glaciologie et de Géophysique de l'Environnement", Grenoble)

francoislouchet38@gmail.com

https://sites.google.com/site/flouchet/

**ABSTRACT**

On the basis of field observations and of a tentative classification of the different types of snow, a modeling of the different possible types of full-depth avalanches is proposed. A new approach for flow and arrest processes is suggested, in terms of healable granular media dynamics.

**INTRODUCTION**

Full-depth avalanches differ from slab avalanches in both release and flow characteristics, and also in season, preferential slope orientation, etc. They are still poorly known, presumably because they are responsible of much less fatalities than those involving slabs laying on weak layers. They cause nevertheless significant damage, and a better knowledge of their triggering processes and of their specific propagation mechanisms should help risk mitigation. The present paper aims at understanding and modeling the physical mechanisms at the origin of their release. After a rapid overview of main full-depth avalanche characteristics, we propose a classification of the different types of snow involved in such avalanches, on the basis of the topological concept of percolation, from which we derive the corresponding triggering mechanisms. Flow and arrest processes are shortly analysed.

**1. OBSERVATIONS:**

Generally speaking, three main types of snow avalanches are usually distinguished:

i) slab avalanches triggered by failure of an underlying weak layer. They are usually named "slab avalanches" for the sake of simplicity, a short cut that may be nevertheless confusing.

ii) loose snow avalanches, triggered by the destabilization of a few snow grains that knock out a couple of other ones, and so on, and resulting in a narrow snow slide flowing from a quasi-punctual starting point.

iii) full-depth avalanches, actually encompassing all other types together, which is the reason why a variety of full-depth avalanche definitions are found in the literature.

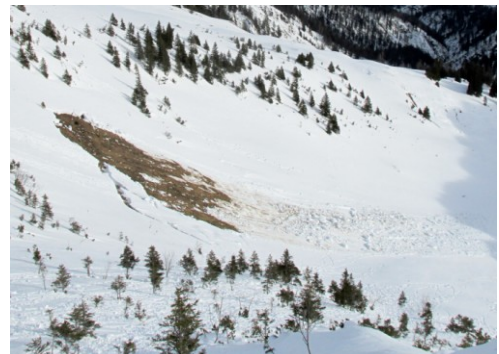

*Fig. 1 Ground avalanche at the Auerspitz*
*08:10, 21 January 2013*
*https://upload.wikimedia.org/wikipedia/commons/6/6e/Grundlawine_auerspitz_201301120.jpg*

In most cases, full-depth avalanches are described as avalanches gliding on a bare ground (fig. 1), but they may also glide on intermediate snow layers if no weak layer is present. Most of them are characterized by a slow flow of high density wet snow (around 500 kg/m$^3$), but they also occasionally involve dry snow (e.g. Col des Aravis, French Alps, december 2008 & 2010). Their widths are usually narrower than those of slab avalanches that may cover a whole hillside. In the northern hemisphere, they often occur in november-december and in march-april, and are less frequent

in january-february. (e.g. winters 2011-2012, 2008-2009, 2007-2008 in the Alps).

A common feature is the absence of weak layer. As a consequence, release is always spontaneous, i.e. full-depth avalanches are not triggered by skiers or by explosives. Starting zone slopes are always steeper than 30°, which means in particular that triggering from a horizontal terrain is impossible, also in relation with the absence of weak layer.

By contrast with slab avalanches, the fact that full-depth avalanches are less frequent in the depths of winter, and that they usually occur on sunny slopes, suggests that melt or rain water plays a significant role. In addition, a strong correlation is observed between full depth avalanche frequency and a temperature close to zero at the ground level, independently of the air temperature during the days before the release [1].

The top of the snow layer starts gliding slowly on the ground surface, whereas the stauchwall may remain immobile, sometimes resulting in spectacular buckling events favoured by warm temperatures.

## 2. DIFFERENT KINDS OF SNOW INVOLVED IN FULL-DEPTH AVALANCHES:

Snow involved in full depth avalanches is usually classified as dry or wet (or damp). A more detailed distinction is sometimes made between wet in the one hand, and sodden, soggy, or waterlogged in the other hand, despite the fact that such qualifiers are not precisely defined in the literature. We propose here definitions of damp snow varieties, that will be shown in the following to correspond to quite different possible triggering mechanisms. They are based on the topological concept of percolation, defined and discussed in Appendix 1.

### 2.1. Dry dense snow:

Dry dense snow results naturally from gradual compaction of light snow, driven by a reduction of ice surface energy, and realized through diffusion of individual water molecules on ice grains surfaces at relatively mild temperatures. Such a process results in a thickening of ice bridges between snow grains, reducing the proportion of air vs ice volume, which significantly increases snow density, and also snow strength by 3 orders of magnitude if density is increased from 200 to 600 kg.m$^3$ [2]. A fairly similar result is also obtained by mechanical compaction, for instance during ski track grooming. This dry dense snow will be labeled "dry snow" hereafter.

### 2.2. Wet snow:

Wet dense snow is formed either by bulk snow thawing at warm temperatures, by rain water infiltration, or both of them. Snow is indeed lighter than ice, due to its significant air content, and even more than water. For this reason, damp snow, in which air has been replaced by water in part or in whole, is always a dense material. In contrast with dry snow (section 2.1), bonds between snow grains have no reason to get stronger. Thawing decreases indeed snow volume fraction, for the benefit of water, resulting in a reduced snow strength. At a given stage of water infiltration, the water volume fraction may exceed the bipercolation threshold (Appendix 1). Snow still percolates through the snow layer, providing mechanical strength to the system, but liquid water also percolates, i. e. is able to find continuous paths through the snow layer, reach the bare ground and lubricate the potential glide surface for the avalanche. In the following, this type of snow will be called "wet snow".

### 2.3. Soggy snow:

An extreme case is that at which water still percolates, but snow grains do not percolate any more, due to an excess of water. At this stage, damp snow suddenly experiences a strong strength discontinuity that changes the medium from a solid state to a granular slurry, due to the fact that the continuous solid skeleton previously made by percolating snow has now disappeared. Such a state will be called "soggy snow".

## 3. RELEASE MECHANISMS

### 3.1 Basal crack destabilization:

A general feature of full-depth avalanches is the absence of weak layer, which eliminates any basal crack initiation by collapse. Artificial triggering (human or by explosives) is therefore almost impossible. In addition, another type of glide surface for the avalanche has to be found in replacement of the weak layer. In the case of slab avalanches, the collapsed weak layer provides indeed a smooth glide plane whose slide resistance is quite low, due to the loose granular character of the crunched weak layer material. By contrast, in the present case, as the weak layer does not exist, the most frequent and obvious glide surface is the ground, if soaked by a sufficient amount of water. This situation may be achieved in different ways, according whether snow is wet, soggy or dry, as discussed in sections 3.2 to 3.4.

Another and fundamental consequence of the absence of weak layer is that we are facing a much simpler problem than in the slab avalanche case, since the only loading component involved in triggering is the shear component parallel to the slope. As a result, the basal crack critical size for unstable crack expansion (in Griffith's sense [3]), now in pure shear, is significantly larger than in mixed shear-compression mode [4], typically several tens of meters instead of a few decimeters, making release more difficult.

The subsequent triggering mechanism is therefore very simple, similar to what was described in early pure-shear models [5]. Griffith's criterion may again be applied (but involving only a pure shear component), in order to determine the critical size at which such flaws become unstable. The shear stress at the interface is given by:

$$\tau = \rho g h_\perp \sin\alpha = \rho g (h\cos\alpha)\sin\alpha = \frac{\rho g h}{2}\sin 2\alpha \quad (1)$$

where $\rho$ is the snow density, $g$ the gravity constant on earth, $h$ the snow depth measured vertically, $h_\perp$ the snow depth measured perpendicular to the ground, and $\alpha$ the slope angle. Assuming that snow falls vertically in average, this shear stress goes through a maximum for a slope angle of $45°$, which is a compromise between a steep slope that increases the shear component of the snow weight along the slope (proportional to *sin α*), and a gentler slope that favors a thicker deposited snow layer (proportional to *cos α*).

In such conditions, Griffith's criterion writes:

$$\tau \sqrt{\pi a_c} = K_{IIc} \quad (2)$$

where $K_{IIc}$ is the snow shear fracture toughness, and $a_c$ is the critical flaw size. Using eq. (1), eq. (2) becomes:

$$\frac{\rho g h}{2}\sin 2\alpha \sqrt{\pi a_c} = K_{IIc} \quad (3)$$

or equivalently:

$$a_c = \frac{4 K_{IIc}^2}{\pi (\rho g h)^2 \sin^2 2\alpha} \quad (4)$$

Eq. (4) shows that, during water percolation through the snow cover, the critical flaw size that destabilizes the avalanche is reached all the more rapidly that $a_c$ is small, i.e. that the snow cover is dense and thick. Rain water contributes in a significant increase of snow density, thus decreasing the critical flaw size and favouring the release process. Eq. (4) also shows that a slope angle close to 45° (that maximizes $\sin^2 2\alpha$) favours triggering. This is derived under the simplifying assumption of a uniform snow depth. Release may obviously take place on gentler slopes due to snow accumulation or particular ground configurations. This finding nevertheless agrees with the fact that full depth avalanches are unlikely to release for slope angles less than 30°.

### 3.2. Wet avalanche release:

We are facing here two percolation problems. The first one is a 3-d percolation through the snow layer, and the second is a 2-d one at the snow / ground interface (appendix 1).

As mentioned above, in the snow layer itself, snow percolates (see definition of "wet snow" in section 2.2), still providing some solid strength to the system, but water also percolates, and is drained through the snow cover down to the ground, favouring snow glide.

The presence of water at the ground level, and a temperature close to zero, are reminiscent of glacier glide kinetics. Glaciers whose temperature at the ground level is negative, called cold glaciers, glide on the glacier bed at low velocities, due to significant friction. By contrast, in warmer glaciers, the temperature at the interface may reach 0°C, resulting in ice melting. The presence of water lubricates glacier glide, significantly increasing glide velocity, and leading in many cases to instabilities and ice falls [6].

The situation in the case of wet avalanches is somewhat similar. Water percolation down to the ground gradually increases its temperature, that eventually reaches the melting point at the ground level. A similar effect can occur due to the geothermal flux, without any contribution of water percolation. As a consequence, water accumulates at places as molten snow flaws, that gradually develop.

At this stage, two different phenomena may now help destabilization of the snow cover, depending on the degree of water percolation at the snow/ground interface, as shown by Faillettaz et al. in the case of glaciers [6]:

i) during early development of molten flaws, the snow layer remains anchored to the ground by a continuous array of frozen snow / ground interfaces, separating isolated molten zones: frozen interfaces percolate at the interface, but molten zones remain isolated from one another. Due to a "water column" effect (height of water drainage paths through the snow cover), molten flaws are pressurized, that help uplifting and uncoupling the snow cover from the ground.

ii) in a second stage, since bipercolation cannot exist at 2 dimensions (appendix 1), If water continues being drained through the snow layer, molten zones grow, and a percolation inversion may take place. Frozen zones become now separated by a continuous array of molten ones. This inversion allows circulation and evacuation of water at the snow / ground interface. Such a mechanism results in some depressurization of molten flaws, but enhances drainage of warmer water through the snow cover. Thus, anchoring zones made of frozen spots readily disappear, considerably increasing lubrication at the bed level.

In addition to this thermodynamical effect, a mechanical phenomenon helps the expansion of molten zones. Such zones behave actually as basal cracks, experiencing the shear stress due to the slope-parallel component of the slab weight. When a molten zone gradually grows due to water circulation, it eventually reaches Griffith's size, becomes unstable, and expands readily to the whole slope. Buckling of the snow cover also participates in flaw growth. The snow cover is now able to glide downslope.

Two additional conditions must however been fulfilled in order to allow long range avalanching: i) opening of a crown crack at the top of the snow layer, and ii) failure of the stauchwall at the bottom.

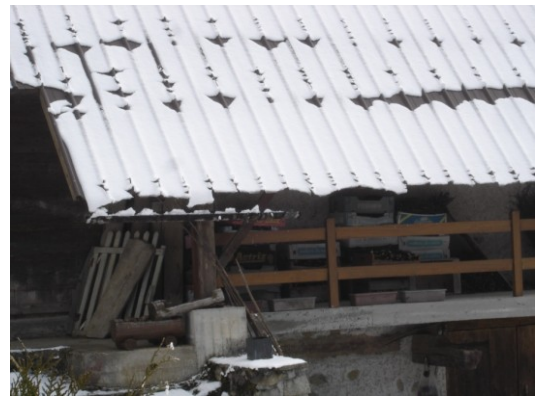

*Fig 2: Incipient wet snow full-depth avalanche on a barn roof. Numerous crown cracks open at every weak place, due to low rupture stresses of wet snow. Note the limited lateral extension of such cracks, related to low yield stresses that favour easy blunting (see text). Some of them may merge (on the right) giving large cracks that may trigger the avalanche.*

Owing to the large amount of water in the snow layer, the tensile rupture stress is low. Crown cracks can open easily as gaps at any weak place, in excellent agreement with observations (fig 2). The shear rupture stress being also reduced in the same proportion, snow may fail by shear from both tips of the crown crack (fig 3), resulting in several narrow avalanches released on the same slope. The snow yield stress being also low, the plastic zone size is large, and crack tips can blunt quite easily (fig. 2, fig. 4). The system now follows the curve in (fig. A2 a) instead of the dashed line.

Rupture becomes ductile and may be delayed, which means that it may occur only for a significantly larger deformation, or may not occur, again in agreement with observations (fig 2).

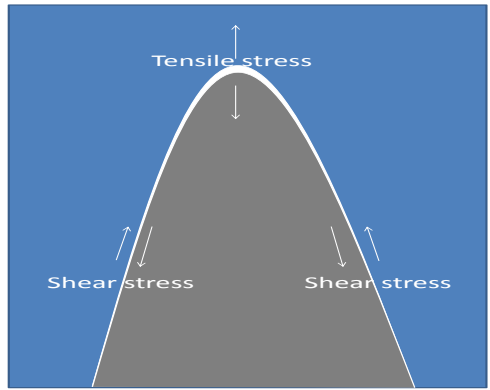

*Fig. 3: Tensile failure at the crown, and lateral failures in shear mode on both sides.*

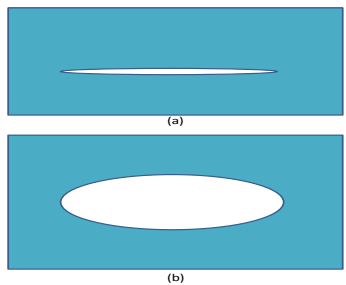

*Fig. 4: (a) acute crack, (b) blunted crack.*

In the case of "classical" slab avalanches, the slab stiffness prevents extensive glide as long as the stauchwall does not fail. In the present case, owing to the ductile deformability of wet snow, the upper part of the snow layer may start deforming by creep, and also sliding downslope, whereas the lower part may not, due for instance to an increased friction or a stauchwall. Such a phenomenon may result in buckling events [7].

### 3.3. Soggy snow avalanche release

At this stage, snow grains do not percolate any more. We deal with some kind of high viscosity granular material, that may deform very slowly at least in the beginning. During such a slow motion, some snow grains inbedded in water may weld together, while bonds between other ones may break. The global slurry behaviour results from a kinetic balance between these two opposite processes. An evolution equation, similar to that first derived in [8], and more extensively in [9] in the case of collapsed weak layer behaviour, can be applied here, showing that if the slowly increasing creep rate reaches a given threshold, the snow would suffer a sudden discontinuous viscosity decrease (bifurcation), and flow downslope more rapidly. Such a mechanism is not incompatible with the presence of blocks of solid snow, often observed in the avalanche flow, provided they are imbedded in the slurry. This type of triggering does not involve any crown crack at least in its traditional definition, as the slurry is already fluid.

### 3.4. Dry snow full depth avalanches:

The most obvious origin of water at the bottom of a dry snow layer is local thawing due to the geothermal flux from a warmer ground (otherwise snow would not be dry!). The mechanism is similar to that of the wet snow case: continued thawing at the bottom of the snow cover gradually increases the sizes of molten flaws, that gradually merge and eventually fulfill Griffith's criterion.

At the same time, the slope-parallel tensile load at the upper rim of the basal crack (i.e. at the crown) raises, thus increasing the stress intensity factor $K = \sigma\sqrt{\pi c}$ at possible snow layer flaws at the crown level, where c is the considered flaw size in the slab itself. Dry snow being an elastic-brittle medium, the situation is fairly similar to that of classical slab avalanches, except that the possibility of weak layer clotting [9] does not exist any more. The system follows the straight dashed line of fig A2 (appendix 2), and eventually reaches the critical value, i.e. the dry snow tensile toughness. A crown crack readily opens and expands rapidly, leading to avalanche release.

### 4. PROPAGATION AND ARREST

Due to a rough glide surface, the release of a wet or soggy snow avalanche readily transforms the snow layer into a water-soaked tumbling granular slurry. Combined with its high density, it propagates at a fairly low velocity. In addition, the flow exhibits significant density and velocity gradients and heterogeneities, making fairly difficult a description using classical Navier Stokes equations for instance. However, as mentioned above, a particular property specific of granular materials is likely to play a key role in both propagation kinetics and arrest mechanism [9]: snow grains collide during the downslope flow of the slurry. They may weld together if their mutual contact time is large enough ("clotting"), allowing $H_2O$ molecules to diffuse along ice grains interfaces, thus thickening bridges between grains. Such a mechanism occurs at low shear strain rates, whereas it is unlikely to occur at larger ones. This problem was mathematically solved in the case of a collapsed weak layer through the derivation of an evolution equation [9], and also applies in the case of full-depth avalanche flow. In agreement with the above qualitative discussion, a critical strain rate threshold is found below which the system made of tumbling individual grains heals by clotting into a solid. This is an example of a bifurcation, at which the material viscosity suffers a drastic and discontinuous increase of several orders of magnitude. The critical shear strain rate is likely to be met first at the basis of the slurry flow (ground level), whose strain rate is reduced at places by ground rugosity, and also at the avalanche tip, due to a diminishing slop angle, that eventually stops the entire flow into a high density solid mass. This is a very simple physical explanation for wet snow avalanche pile-ups, as compared to classical fluid dynamics models that are unable to account for such a flow and arrest mechanism, except using complicated multi-parameter phenomenological models. Introduction of such a threshold in these flow simulations should be of interest.

## 5. SUMMARY AND CONCLUSION

By contrast with slab avalanches that are triggered by weak layer collapse and propagate on the weak layer surface, full-depth avalanches glide (by definition) on the bare ground. As a consequence, they can hardly be triggered by skiers or explosives. Water plays a key role in release mechanisms. Its origin (snow thawing at the bottom of the snow layer, bulk snow melting or rain water drainage) determines the type of avalanche. Such types were discussed here in terms of percolation. In the case of "soggy snow", water percolates, but snow does not any more, that results in flow of a snow-water slurry. In the case of "wet snow", water / snow bipercolation in the snow layer leads to development of flaws of molten snow at the ground level, that may in turn percolate and trigger the avalanche, that also turns into a water/snow slurry flow due to ground roughness. Dry snow avalanche release is more similar to that of slab avalanches, with the key difference that weak layers do not exist in this case. In all cases, triggering is controlled by shear stresses. Flow and arrest characteristics are essentially ruled by dynamics of healable granular media.

## APPENDIX 1: PERCOLATION

The topological concept of percolation, reminiscent of coffee making, is quite important in various physical problems, particularly in yielding and fracture mechanisms, and more specifically in the case of full-depth avalanches. It can be illustrated in a very simple way, as follows.

Let us consider a coffee filter filled with coffee powder, and let us pour water on it. Three different situations may be considered:

i) If the powder is firmly crammed down into the filter, hot water poured on it would stay at the surface, unable to go through the powder. In this case, it is possible to find connected paths made of coffee grains in mutual contact that go through the entire system. Coffee grains are said to "percolate" through the system, but water does not, and coffee making becomes a complicated task.

ii) If the coffee powder is less tightly packed, water would be able to find a way through the powder, down to the coffee pot. It percolates through the system, but the coffee powder still percolates, which prevents its mechanical collapse. This is called "bi-percolation". It is worth noting that bi-percolation cannot occur in 2-dimensional systems: in a 3-d coffee maker, water is able to use the $3^d$ dimension to bypass a continuous coffee obstacle, which is impossible in 2-d.

iii) if too much water is poured at the same time on the coffee powder, the mixture turns into a fluid. Water percolates, but coffee does not any more. This is the situation found in " turkish coffee" making.

Physical and more specifically mechanical properties of random media drastically change at so-called percolation thresholds, where isolated clusters may become connected into a theoretically infinite network, or conversely. Water may percolate in snow, and bi-percolation or "mono-percolation" may result in different full-depth snow avalanche types.

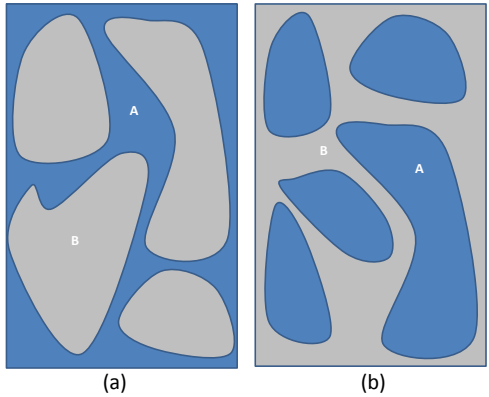

(a)         (b)

*Fig. A1: Percolation: (a) A percolates in B; (b) B percolates in (A). A and B can percolate simultaneously only in 3-dimensions.*

**APPENDIX 2: THE BRITTLE TO DUCTILE TRANSITION**

Anybody knows that a block of butter left by the window in summer splashes as a lump of mud if dropped on the ground, whereas it breaks into pieces if just taken out of the freezer. Butter is brittle at low temperatures, and ductile at warmer ones. This phenomenon is quite general, and is known as the brittle - ductile transition.

The explanation can be given in a simple way on the basis of a combination of Griffith's criterion and of the plastic behaviour of the material. Butter from the window is plastic, i.e. it can change its shape under stress in an irreversible way. By contrast, butter from the freezer is elastic: it deforms under stress (though in a lesser extent than a rubber tape), but recovers its previous shape upon stress reduction. Under stress, it may either splash or elongate (depending on the compressive or tensile stress character) in the former case (it is ductile), whereas it breaks off suddenly in the latter one (it is brittle).

Such different behaviours can be illustrated using stress concentration vs time diagrams. Let us consider a brittle material, and assume for the sake of simplicity that the stress intensity factor $K = \sigma \sqrt{\pi a}$ increases linearly with time, as schematized by the straight line with a slope $\dot{K}$ (fig. A2). The material fails at a time $t_{RE}$ when the straight line of slope $\dot{K}$ intersects the horizontal line $K = K_c$ (Griffith's criterion).

Let us now consider a plastic material. As a first approximation, the plastic character means that if stress exceeds a so-called "yield stress", the material deforms plastically, i.e. in a slow and irreversible way. The yield stress of steel is significantly larger than that of "butter from the window", and requires temperatures by far higher, but the physics of deformation are fairly similar.

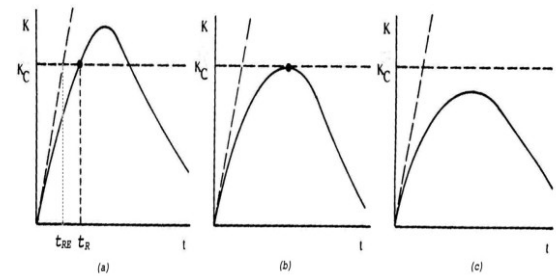

*Fig. A2. Stress intensity factor K vs time diagrams. (a) Pure elastic case (dashed line) and brittle plastic case (solid curve) for which the failure time is delayed ($t_R > t_{RE}$). (b) Bifurcation: Brittle to ductile transition. (c) Ductile case: K never reaches the critical $K_c$ value: the material tears apart instead of breaking off.*

Let us now focus at what happens in the vicinity of the flaw in a stressed plastic material. Crack tips are an example of "stress concentrators", which means that the local stress at the crack tip itself tends to infinity if the tip is acute. It remains large in the vicinity of the tip, and gradually decreases at larger distances. The zone within which the stress exceeds the yield stress is called the plastic zone. It is obvious that the plastic zone size increases for decreasing yield stresses, or in other words as temperature is increased for a given material. Plastic activity in the plastic zone results in a blunting of the crack tip (fig. 4), which in turn reduces the stress concentration. As a consequence, instead of following the dashed straight line of fig. A2 a, the stress intensity factor in the case of a moderately plastic material grows up more slowly, and the critical value $K_c$ is reached after a longer time $t_R > t_{RE}$ (fig. A2 a). If the material becomes more and more plastic (temperature increase, or larger water concentration in wet snow), the *K(t)* curve goes through a bifurcation corresponding to the brittle-ductile transition (fig. A2 b), and eventually becomes totally ductile (fig. A2 c) since it cannot reach the $K_c$ value any more.

**ACKNOWLEDGEMENTS**:

The author thanks Jérôme Faillettaz (Zürich University) for quite interesting discussions on glacier dynamics, and Alain Duclos & Greg Coubat, (Data-Avalanche Association), for fruitful remarks on the manuscript

.